\documentclass[aps,twocolumn,superscriptaddress]{revtex4-2}
\usepackage{natbib}
\usepackage{graphicx}
\usepackage{amsmath}
\usepackage{amssymb}
\usepackage{amsmath}
\usepackage[colorlinks,urlcolor=blue,citecolor=blue,linkcolor=blue]{hyperref}
\usepackage{url}

\begin{document}

\title{Superconductor-polariton non-dissipative drag in optical microcavity}

\author{Azat F. Aminov}
\email{afaminov@hse.ru}
\affiliation{National Research University Higher School of Economics, 109028 Moscow, Russia}

\author{Alexey A. Sokolik}
\email{asokolik@hse.ru}
\affiliation{Institute for Spectroscopy, Russian Academy of Sciences, 142190 Troitsk, Moscow, Russia}
\affiliation{National Research University Higher School of Economics, 109028 Moscow, Russia}

\author{Yurii E. Lozovik}
\affiliation{Institute for Spectroscopy, Russian Academy of Sciences, 142190 Troitsk, Moscow, Russia}
\affiliation{National Research University Higher School of Economics, 109028 Moscow, Russia}

\begin{abstract}
We consider non-dissipative drag between Bose-condensed exciton polaritons in optical microcavity and embedded superconductors. This effect consists in induction of a non-dissipative electric current in the superconductor by motion of polariton Bose condensate due to electron-polariton interaction, or vice versa. Using many-body theory, we calculate the drag density, characterizing magnitude of this effect, with taking into account dynamical screening of the interaction. Hoping to diminish the interaction screening and microcavity photon absorption, we consider atomically-thin superconductors (both conventional s-wave and copper-oxide d-wave) of planar and nanoribbon shapes. Our estimates show that in realistic conditions the drag effect could be rather weak but observable in accurate experiments in the case of dipolar interlayer excitons in transition metal dichalcogenide bilayers. Use of spatially direct excitons, semiconductor quantum wells as the host for excitons, or thin films of bulk metallic superconductors considerably lowers the drag density.
\end{abstract}

\maketitle

\section{Introduction}

Non-dissipative drag was initially predicted for a mixture of superfluid $^{3}$He and $^{4}$He \cite{AB}, and then studied theoretically in other systems: ultracold Bose-condensed atomic gases \cite{Fil2005}, two interacting superconductors \cite{Duan1993}, and superfluid proton-neutron mixtures in neutron stars \cite{Alpar1984}. Unlike the Coulomb drag in heterostructures \cite{Narozhny2016} or electron-polariton systems \cite{Berman2016}, this effect is non-dissipative: in a mixture of two superfluids there appears the cross-coupling between velocities of superfluid components and superfluid currents of two species with the coefficient called drag density \cite{AB}.

Several experimental methods to detect the non-dissipative drag were proposed over the years. The most straightforward one implies measuring the current, induced in one specie of the mixture in response to the current in the other specie \cite{AB,Fil2005}. Another method is to measure the spin susceptibility and the speed of sound of the spin mode (antiphase oscillation of both components) \cite{Romito2020,Nespolo2018}, since these quantities can be straightforwardly related to the drag density. Moreover, it was shown that the non-dissipative drag affects the stability criteria of Bose-Bose superfluid mixtures \cite{Nespolo2018}. In extended 2D systems, where the Berezinskii-Kosterlitz-Thouless transition is expected, the large drag density can destroy the superfluidity \cite{Karle2019}. Other detection methods include, for example, measuring of the magnetic fields, induced by drag currents in superconductors \cite{Duan1993}, and interferometry of ultracold atomic gases \cite{Khalid2021}. Despite predictions of the non-dissipative drag in various systems, it is still elusive in the experiments and has not been directly confirmed.

Recent advances in condensed matter physics, especially in two-dimensional superconductivity \cite{Uchihashi2017} and electromagnetic microcavity designs \cite{QFl}, may open new possibilities for detection of the drag effect. One of the involved superfluid subsystems may be a Bose-Einstein condensate (BEC) of excitonic polaritons, which was realized in a lot of materials, such as bulk or nanostructured semiconductors (CdTe, GaAs, GaN etc.), atomically thin transition metal dichalcogenides (TMDCs), and organic compounds. The discovery of graphene paved the way for two-dimensional superconductors (FeSe, magic-angle twisted bilayer graphene, TMDCs, copper oxide layers etc.), which can be used as the second subsystem. The resulting hybrid superconductor-polariton system provides opportunities to study Bose-Fermi collective effects, which are starting to draw attention \cite{Sven2021,Cotlet2016}.

In this paper we investigate the non-dissipative drag between two systems of different nature: Bose condensate of interlayer excitonic polaritons, arising in TMDC bilayers embedded in Fabry-Perot optical microcavity, and Cooper-pair condensate of a superconductor. Unlike in our previous paper \cite{Aminov_2022}, where the superconducting system was assumed to be a two-dimensional (2D) electron gas with the Cooper pairing induced by the polariton-BEC mediated electron attraction \cite{Cotlet2016,Laussy2010}, here we consider the pre-existing superconductors of extreme thinness (NbSe$_2$, FeSe, YBCO, LSCO, and $\mathrm{Nb}_{1-x}\mathrm{Ti}_x\mathrm{N}$). The choice of these materials is motivated by intention to reduce both the absorption of microcavity photons and the screening of electron-polariton interaction by a superconductor. To reduce the influence of superconductor even more, we consider not only 2D thin-film, but also one-dimensional (1D) ribbon- and wire-like superconductor geometries. 

\begin{figure*}[t]
\begin{center}
\includegraphics[width=0.75\textwidth]{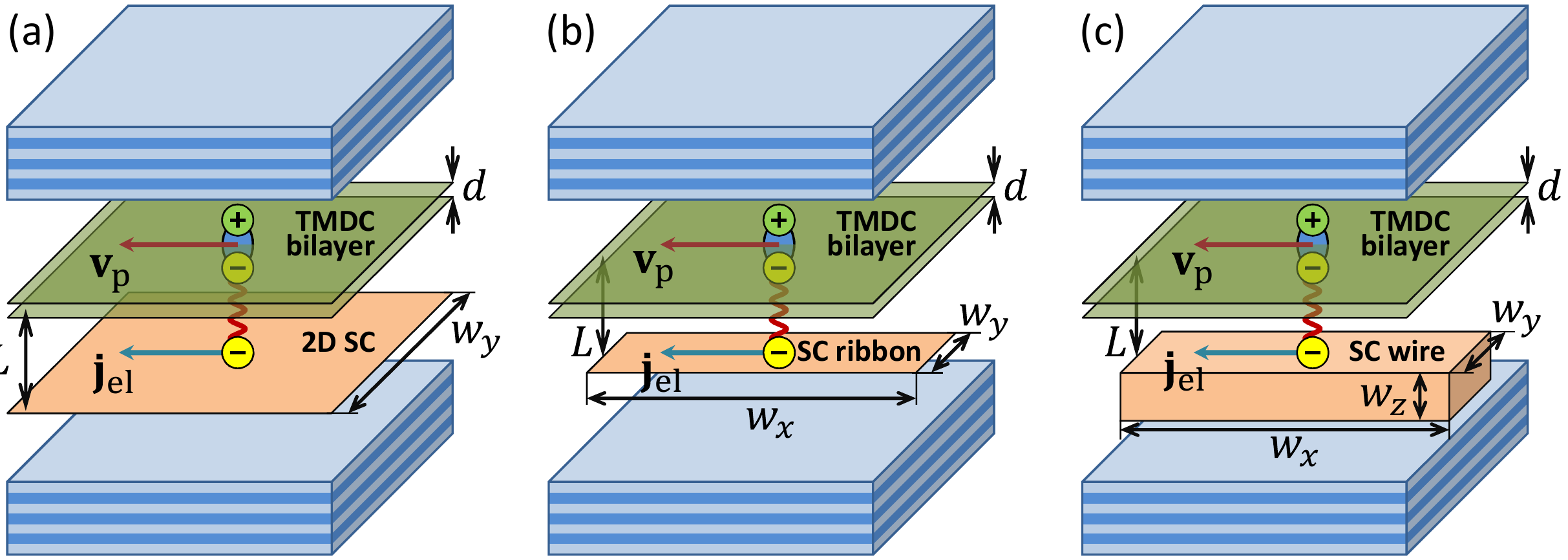}
\end{center}
\caption{\label{Fig1}System schematic: Bose-condensed indirect polaritons in TMDC bilayer embedded into Fabry-Perot optical microcavity interact with (a) 2D atomically thin superconductor (SC), (b) 1D atomically thin superconducting ribbon, or (c) 1D wire fabricated from bulk superconductor. Due to electron-exciton interaction, the superfluid drag appears when the motion of exciton-polariton BEC with the velocity $\mathbf{v}_\mathrm{p}$ entrains Cooper pair condensate giving rise to the non-dissipative current of electrons $\mathbf{j}_\mathrm{el}$.}
\end{figure*}

The overview of the superconducting materials taken into consideration is given in Sec.~\ref{Sec2}. As described in Sec.~\ref{Sec3}, we introduce the drag density which unambiguously characterizes magnitude of the non-dissipative drag even if the effective mass of electrons in a superconductor is undefined, as happens for high-temperature copper-oxide superconductors with essentially anisotropic electron dispersions and Fermi surfaces. Relating the drag density to the correlation function of currents, we perform many-body calculations of this quantity with taking into account dynamical screening of electron-polariton interaction by density responses of polariton and electron subsystems, generally anisotropic electron dispersion and superconducting gap, and different superconductor geometries.

The calculation results for the drag density in realistic conditions are presented in Sec.~\ref{Sec4}. We show that nanoribbon geometry allows to achieve the same order of drag density as with atomically thin 2D superconductors, but at larger distances between electronic and excitonic layers. In the case of direct excitons in TMDC monolayer, or with nanowires made of bulk superconductors, the drag density turns out to be much lower. Sec.~\ref{Sec5} is devoted to discussion of possible experimental observation of the effect, and Appendices~\ref{Appendix_A} and \ref{Appendix_B} present details of calculations.

\section{System overview}\label{Sec2}

The main issue with embedding a superconductor into an optical microcavity is to avoid degradation of the quality factor due to light absorption by the superconductor film. To achieve it, one can use atomically thin 2D superconductors or refuse from the plane-parallel geometry in order to reduce spatial overlap between the superconductor and microcavity electromagnetic mode. The latter scenario may include placement of a superconductor around the microcavity \cite{Skopelitis,Sedov2020} or in a shape of narrow nanowire.

Three kinds of system schematics we propose are shown in Fig.~\ref{Fig1}. The first one [Fig.~\ref{Fig1}(a)] includes 2D atomically thin superconductor with low absorption and transverse width $w_y$ much larger than the out-of-plane distance $L$ between the superconductor and the excitonic layer. In the second case  [Fig.~\ref{Fig1}(b)], the atomically thin superconductor has a shape of 1D nanoribbon with the width $w_y\ll L$, which allows to reduce the light absorption and the screening of electron-polariton interaction by the superconductor. The third case [Fig.~\ref{Fig1}(c)] deals with the 1D nanowire made of bulk metallic superconductor with the width $w_y$ and thickness $w_z$, both small with respect to $L$. In all cases, the dipole spatially indirect excitons in TMDC bilayer are shown in the figure, although the case of spatially direct excitons in a monolayer will also be studied in the geometry of Fig.~\ref{Fig1}(a).

\begin{table*}[t]
\centering
\begin{tabular}{llllll}
\hline\hline
& NbSe$_{2}$ & FeSe & YBCO & LSCO & $\mathrm{Nb}_{1-x}\mathrm{Ti}_x\mathrm{N}$\\
\hline
\multicolumn{6}{c}{Optical parameters} \\
\hline
Extinction & 0.5 \cite{Lin_2019} & 0.14 & --- & --- & 2.5 \\
$\alpha$ ($10^{5}$ cm$^{-1}$) & 0.9 & 0.25 \cite{Ouertani_2021} & 0.77 \cite{Samovarov_2003} & 0.16 \cite{Schwartz_2000} & 4.5 \\
$h$ & 2.2 nm& 0.55 nm & 1.17 nm & 1.32 nm & 3 nm \\
$\alpha h$ & 0.02 & 0.0014 & 0.009 & 0.002 & 0.14 \\
\hline
\multicolumn{6}{c}{Electronic parameters} \\
\hline
$T_{\mathrm{c}}$ & 4.7 K & 50 K & 30 K & 30 K & 16 K \\
$\Delta$ & 0.7 meV & 17 meV & 20 meV & 20 meV & 4 meV \\
$n_\mathrm{el}$ & $10^{15}$ cm$^{-2}$ & $10^{14}$ cm$^{-2}$  & $10^{14}$ cm$^{-2}$ & $5 \times 10^{13}$ cm$^{-2}$ & $10^{22}$ cm$^{-3}$ \\
$m_\mathrm{el}^*/m_0$ & 1 & 2.7 & --- & --- & 1 \\
$j_\mathrm{c}^\mathrm{3D}$& $2\times10^7\,\mbox{A/cm}^2$ & $3\times 10^5\,\mbox{A/cm}^2$ & $10^6\,\mbox{A/cm}^2$ & --- & $4\times10^6\,\mbox{A/cm}^2$ \\
\hline\hline
\end{tabular}
\caption{Extinction and attenuation $\alpha$ coefficients, thicknesses $h$, and absorption coefficients $\alpha h$ at the wavelength 700~nm for the superconductors we consider along with their electronic parameters: critical temperature $T_\mathrm{c}$, zero-temperature superconducting gap $\Delta$, density $n_\mathrm{el}$ and effective mass $m_\mathrm{el}^*$ of charge carriers, and three-dimensional (3D) critical current density $j_\mathrm{c}^\mathrm{3D}$. The parameters for NbSe$_{2}$ are taken from \cite{Lin_2019,Khestanova_2018,H_rhold_2023,He_2020}, for FeSe from \cite{Ouertani_2021, Liu_2012, Sun_2014}, for YBCO and LSCO from \cite{Samovarov_2003, Schwartz_2000, Koblischka_Veneva_2019}, and for $\mathrm{Nb}_{1-x}\mathrm{Ti}_x\mathrm{N}$ from \cite{Holzman_2019, Sidorova_2021} }
\label{table:1}
\end{table*}

As candidate materials for our calculations, we consider the following atomically thin superconductors: NbSe$_2$ \cite{Khestanova_2018,Lin_2019}, FeSe \cite{Liu_2012}, as well as copper-oxide high-temperature superconductors YBCO \cite{Samovarov_2003,Bari_i__2013} and LSCO \cite{Schwartz_2000}. For 1D wires, we consider relatively thick niobium-based 3D superconductors $\mathrm{Nb}_{1-x}\mathrm{Ti}_x\mathrm{N}$ \cite{Holzman_2019}. The parameters of all superconducting materials are listed in Table~\ref{table:1}. The attenuation coefficients $\alpha$ of 2D superconductors are low enough for the integral absorption coefficient $\alpha h$, with taking into account their effective optical thickness $h$, to be less than 0.01, so we might hope that the microcavity quality will not degrade considerably, especially when the superconductor layer does not overlap its cross section completely.

Besides the superconducting materials presented in the Table \ref{table:1}, there are other superconductors with low absorption. For instance, so-called transparent superconductors, like Li$_{1-x}$NbO$_{2}$ with an attenuation around $0.2\times 10^5\,\mbox{cm}^{-1}$ at the wavelength 700~nm \cite{Soma_2020}. However, the thickness of such superconductors in experiments is usually of the order of 100~nm, so their embedding into microcavity may be a challenging task.

Some remarks about the properties of FeSe and NbSe$_2$ listed in Table~\ref{table:1} should be made. The FeSe monolayer can be grown on various substrates, like SrTiO$_3$, BaTiO$_3$, graphene, TiO$_{2}$ (see \cite{Nekrasov_2018} and references therein). We consider the SrTiO$_3$ substrate, which is almost transparent and provides the highest critical temperature. Although the Fermi surface of FeSe contains additional hole pocket \cite{Liu_2012}, with this substrate it is located well beneath (at 80 meV) the Fermi level and can be disregarded. NbSe$_2$ has three hole pockets, namely K, K$^{\prime}$ and $\Gamma$ \cite{H_rhold_2023}. While there are differences between them, the total answer does not change drastically on the exact form of these pockets, so we model NbSe$_{2}$ as three-valley material. There is an on-going debate on the exact form the superconducting gap and its momentum dependence \cite{Sanna_2022, Khestanova_2018}, and for simplicity we assume the same isotropic gap $\Delta=0.7\,\mbox{meV}$ \cite{Khestanova_2018} in all three valleys.

\section{Theory}\label{Sec3}

Non-dissipative drag is characterized by the tensor of electron-polariton current response
\begin{equation}
\chi_{\alpha\beta}(\mathbf{q},i\Omega)=-\frac1A\int\limits_0^{1/T}d\tau\:e^{i\Omega\tau}\left\langle T_\tau j_{\mathbf{q}\alpha}^{\mathrm{el}}(\tau)j^{\mathrm{p}}_{-\mathbf{q}\beta}(0)\right\rangle.\label{chi1}
\end{equation}
For 2D superconductor $\alpha,\beta=x,y$ and $A$ is the area $w_xw_y$ of the electron layer, while for 1D superconducting ribbon or wire $\alpha=x$, $\beta=x,y$, and $A=w_x$ is its length. In the long-wavelength limit $\mathbf{q}\rightarrow0$, the operators of current density Fourier harmonics $\mathbf{j}^{\mathrm{el}}_{\mathbf{q}\rightarrow0}=\hbar^{-1}\sum_{\mathbf{k}s} (d\epsilon^\mathrm{el}_\mathbf{k}/d\mathbf{k}) a^\dag_{\mathbf{k}s}a_{\mathbf{k}s}$, $\mathbf{j}^{\mathrm{p}}_{\mathbf{q}\rightarrow0}=\hbar^{-1}\sum_\mathbf{k}(d\epsilon^\mathrm{p}_\mathbf{k}/d\mathbf{k}) b^\dag_\mathbf{k}b_\mathbf{k}$ are given in terms of electron $\epsilon^\mathrm{el}_\mathbf{k}$ and polariton $\epsilon^\mathrm{p}_\mathbf{k}$ dispersions as well as creation and destruction operators of electrons $a^\dag_{\mathbf{k}s}$, $a_{\mathbf{k}s}$ and polaritons $b^\dag_{\mathbf{k}}$, $b_{\mathbf{k}}$ with the momentum $\mathbf{k}$; $s$ is the electron spin and valley index.

The static long-wavelength limit 
\begin{equation}
\chi_\mathrm{T}=\lim_{q_x\rightarrow0}\lim_{q_y\rightarrow0}\chi_{xx}(\mathbf{q},i\Omega=0)\label{chi_T}
\end{equation}
of the transverse part of the tensor (\ref{chi1}) determines response of the homogeneous electron current density $\mathbf{j}_\mathrm{el}$ on the gradient of phase $\varphi_\mathrm{p}$ of the polaritonic BEC:
\begin{equation}
\mathbf{j}_\mathrm{el}=\chi_\mathrm{T}\hbar\nabla\varphi_\mathrm{p}.
\end{equation}
Note that the true superfluidity of polaritons should be not necessary for existence of the non-dissipative drag, since we need only BEC with the long-range or quasi long-range order and nonzero gradient $\nabla\varphi_\mathrm{p}$ of the order parameter. The distinction between BEC and superfluidity can be essential for driven-dissipative exciton-polariton systems \cite{Szymaska2006}.

In the case where the polaritons have well-defined mass, $\epsilon^\mathrm{p}_\mathbf{k}=\hbar^2k^2/2m_\mathrm{p}$, at characteristic velocities of their condensate motion, we can define the polariton condensate velocity $\mathbf{v}_\mathrm{p}=\hbar\nabla\varphi_\mathrm{p}/m_\mathrm{p}$ and the drag density $n_\mathrm{dr}$:
\begin{equation}
\mathbf{j}_\mathrm{el}=n_\mathrm{dr}\mathbf{v}_\mathrm{p},\qquad n_\mathrm{dr}=m_\mathrm{p}\chi_\mathrm{T}.\label{n_dr}
\end{equation}
When the electron effective mass is also well defined, $\epsilon^\mathrm{el}_\mathbf{k}=\hbar^2k^2/2m^*_\mathrm{el}$, we can introduce the drag-induced mass current density of electrons $\mathbf{g}_\mathrm{el}=m_\mathrm{el}^*\mathbf{j}_\mathrm{el}$ and the drag mass density $\rho_\mathrm{dr}$:
\begin{equation}
\mathbf{g}_\mathrm{el}=\rho_\mathrm{dr}\mathbf{v}_\mathrm{p},\qquad \rho_\mathrm{dr}=m_\mathrm{el}^* n_\mathrm{dr}=m_\mathrm{el}^*m_\mathrm{p}\chi_\mathrm{T}.
\end{equation}
In our preceding paper \cite{Aminov_2022} we calculated $\rho_\mathrm{dr}$ implying quadratic approximations for both electron and polariton dispersions. Here we go beyond this approximation in order to include into our analysis the copper-oxide superconductors YBCO and LSCO with manifestly non-quadratic electron dispersions.

Our many-body calculations of the current response (\ref{chi1}) are similar to those in Ref.~\cite{Aminov_2022}, although with several modifications aimed on taking into account both 2D and 1D geometries of the electron system, general forms of electron and polariton dispersions, and both s- and d-wave superconducting gaps. In the second order in the screened electron-polariton interlayer interaction, Eq.~(\ref{chi1}) takes the form
\begin{align}
\chi_{\alpha\beta}(0,0)=-\frac{T}{2S}\sum_{\mathbf{q}\omega_n}\mathfrak{F}_\alpha^\mathrm{el}(\mathbf{q},i\omega_{n})\mathfrak{F}_\beta^\mathrm{p}(\mathbf{q},i\omega_{n})\nonumber\\
\times\left|\tilde{V}^\mathrm{el-p}_{\mathrm{scr}}(q,i\omega_{n})\right|^{2},\label{chi2}
\end{align}
where $\omega_n=2\pi Tn$ are bosonic Matsubara frequencies, and the vector nonlinear rectification functions of electron $\boldsymbol{\mathfrak{F}}^\mathrm{el}$  and polariton $\boldsymbol{\mathfrak{F}}^\mathrm{p}$ subsystems are defined and calculated in Appendix~\ref{Appendix_A}. The key differences between our present calculation and Ref.~\cite{Aminov_2022} are the following: (a) we take into account generally non-quadratic electron $\epsilon_\mathbf{k}^\mathrm{el}$ and polariton $\epsilon_\mathbf{k}^\mathrm{p}$ dispersions by replacing $\mathbf{k}/m_\mathrm{el}^*$ and $\mathbf{k}/m_\mathrm{p}$ with the corresponding group velocities $\hbar^{-1}d\epsilon_\mathbf{k}^\mathrm{el}/d\mathbf{k}$ and $\hbar^{-1}d\epsilon_\mathbf{k}^\mathrm{p}/d\mathbf{k}$; (b) for 1D superconducting ribbons and wires, we take into account that electron current $\mathbf{j}^\mathrm{el}$, rectification function $\boldsymbol{\mathfrak{F}}^\mathrm{el}$, and interlayer transferred momentum $\mathbf{q}$ are 1D vectors directed along the $x$ axis; (c) for d-wave superconductors YBCO and LSCO we take into account anisotropy of the gap.

The partially screened interlayer electron-polariton interaction
\begin{equation}
\tilde{V}_\mathrm{scr}^\mathrm{el-p}(q,i\omega_n)=\frac{V^\mathrm{el-p}(q)}{\tilde\varepsilon(q,i\omega_n)}\label{V_scr}
\end{equation}
entering Eq.~(\ref{chi2}) is the electron-polariton interaction $V^\mathrm{el-p}(q)$ screened by random-phase approximation diagrams with density responses of both electron and polariton systems. The partial character of the screening consists in neglecting the diagrams responsible for the screening by polaritons at the polaritonic end of the interaction line. This kind of screening is already taken into account in the polariton rectification function (\ref{F_p2}), so it should be neglected here to avoid double counting of diagrams \cite{Boev2019,Aminov_2022}.

The bare electron-polariton interaction $V^\mathrm{el-p}(q)$, which is screened only by surrounding dielectrics, was described in Ref.~\cite{Aminov_2022} and differs in the cases of spatially indirect and direct excitons. In the latter case the interaction is much weaker due to absence of excitonic persistent dipole moment, although this setup does not require the bilayer system to host excitons. In Appendix~\ref{Appendix_B} we describe how the dielectric function $\tilde\varepsilon(q,i\omega_n)$ is calculated for both 2D and 1D (ribbon or wire) superconductor geometries. In the latter case we need to take into account 1D density response and 1D electron-electron Coulomb interaction in the electron subsystem. Reduced density response of the superconductor caused by its small transverse $w_y$ and vertical $w_z$ dimensions makes the electron-polariton interaction (\ref{V_scr}) stronger in the 1D geometry than in the 2D plane-parallel geometry. For the YBCO and LSCO copper-oxide superconductors we calculate the density response numerically with taking into account the anisotropies of both electron dispersion and d-wave energy gap.

\section{Calculation results}\label{Sec4}

\begin{figure}[t]
\centering
\begin{center}
\includegraphics[width=\columnwidth]{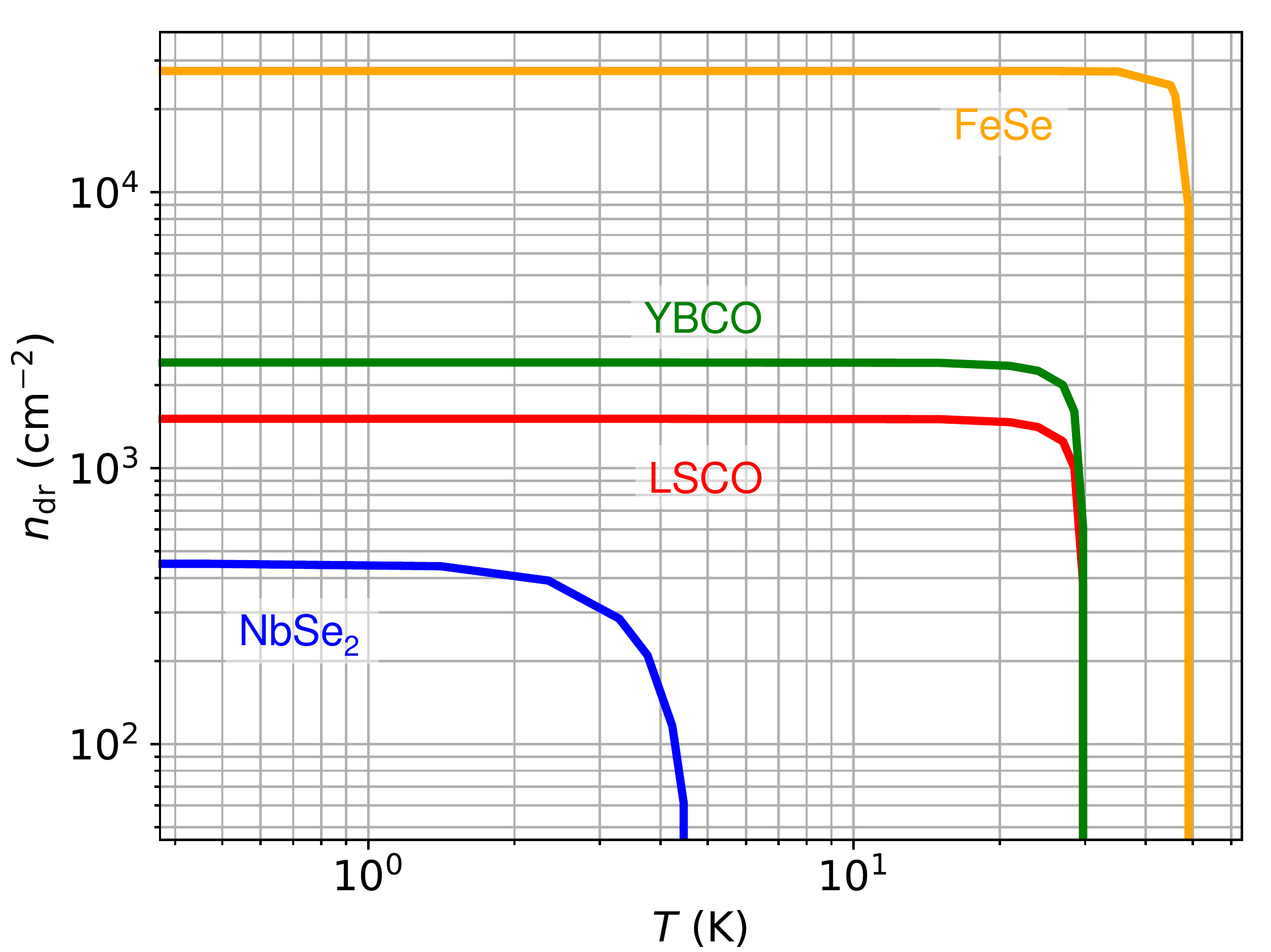}
\end{center}
\caption{\label{fig:T}Drag density $n_\mathrm{dr}$ between Bose-condensed indirect polaritons and 2D superconductors as function of temperature. The Rabi splitting is $\Omega_{\mathrm{R}} = 20\,\mbox{meV}$ and the distance between superconductor and indirect excitons is $L=10\,\mbox{nm}$.}
\end{figure}

We carried out numerical calculations of the drag density $n_\mathrm{dr}$ according to Eqs.~(\ref{chi_T}), (\ref{n_dr}), and (\ref{chi2}) under realistic conditions. The parameters of superconductors are listed in Table~\ref{table:1}. The parameters of polariton subsystem are close to those used in Ref.~\cite{Aminov_2022}. For optical microcavity we take the bare photonic mass $m_\mathrm{c}=5\times10^{-5}\,m_0$ ($m_0$ is the free electron mass) and dielectric constant $\varepsilon_\mathrm{env}=7$. Spatially indirect excitons in TMDC bilayers are considered with the electron and hole effective masses $0.5m_0$, Bohr radius $4\,\mbox{nm}$, vertical electron-hole separation $d=4\,\mbox{nm}$, and exciton-exciton contact interaction $g^\mathrm{x-x}=0.1\,\mu\mbox{eV}\cdot\mu\mbox{m}^2$. Excitons hybridize with the microcavity photons thanks to the Rabi splitting $\Omega_\mathrm{R}=5-30\,\mbox{meV}$ to form Bose-condensed polaritons with the density $n_\mathrm{p}=10^{12}\,\mbox{cm}^{-2}$.

First we analyze the temperature dependence of the drag density (\ref{n_dr}) shown in Fig.~\ref{fig:T} for the atomically thin 2D superconductors $\mathrm{NbSe}_2$, FeSe, YBCO, LSCO. Here and henceforth we assume that the critical temperature of the polariton Bose condensation is much higher than superconductor critical temperatures $T_\mathrm{c}$ and thus neglect the thermal depletion of the condensate. The figure demonstrates that in each superconductor the drag density $n_\mathrm{dr}$ decreases with increasing temperature and vanishes at $T=T_\mathrm{c}$ mainly due to the temperature dependence of the superconducting gap. It can be noted that $n_\mathrm{dr}$ is more robust versus temperature in the superconductors with large superconducting gaps (FeSe, LSCO, and YBCO) rather than in small-gap superconductors (NbSe$_2$). However, a large gap by itself is not essential for high $n_\mathrm{dr}$ at $T\ll T_\mathrm{c}$; NbSe$_2$ shows inferior results in Fig.~\ref{fig:T} mainly due to the strong screening caused by its high electron density (see Table~\ref{table:1}). Since $n_\mathrm{dr}(T)$ decreases rather slowly at $T\lesssim T_\mathrm{c}$, in the following we will plot $n_\mathrm{dr}$ at $T=0$.

\begin{figure}[t]
\centering
\begin{center}
\includegraphics[width=\columnwidth]{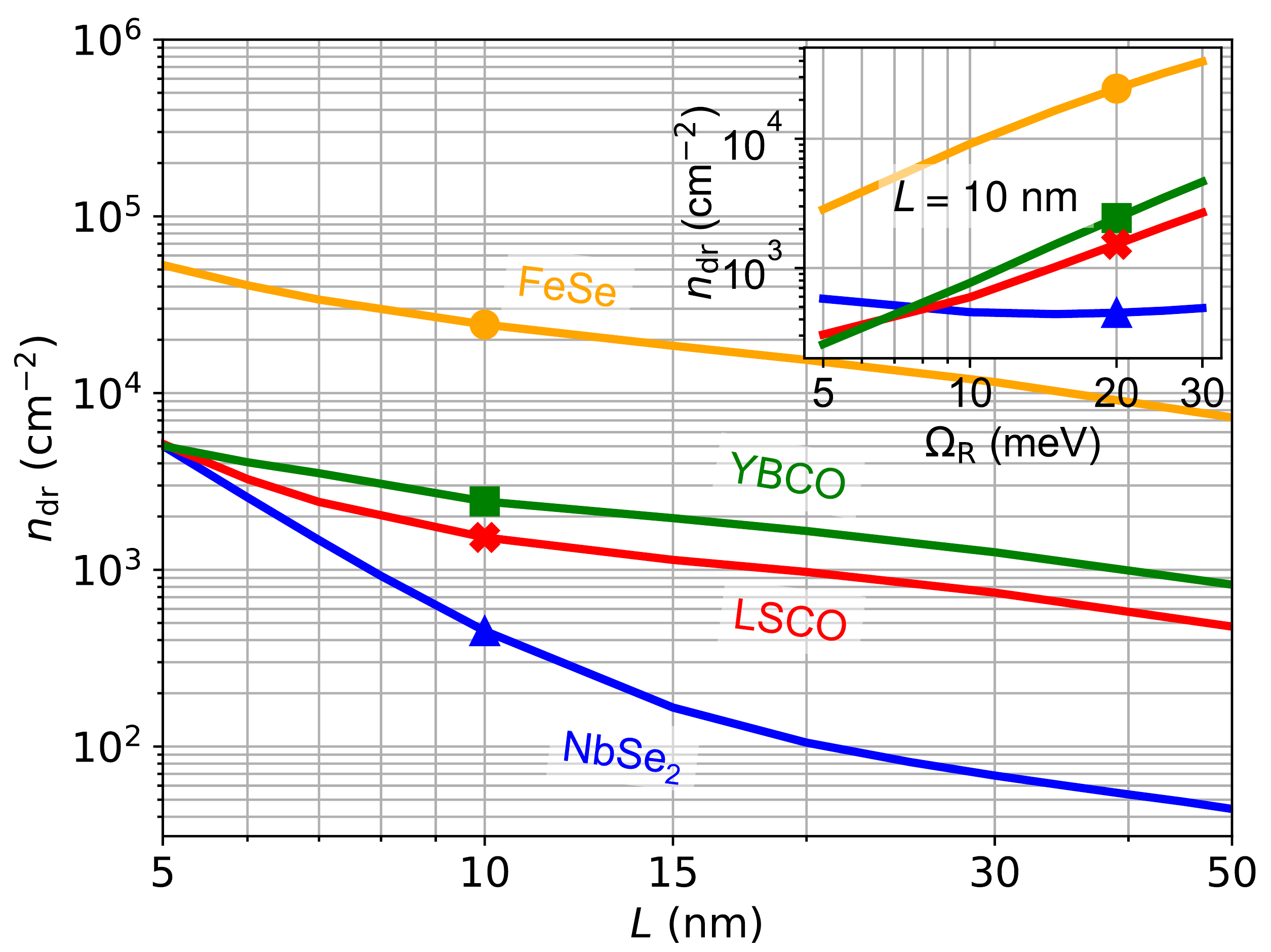}
\end{center}
\caption{\label{fig:p}
Drag density $n_\mathrm{dr}$ as function of interlayer distance $L$ between Bose-condensed indirect polaritons and 2D superconductors. The Rabi splitting on the main plot is $\Omega_{\mathrm{R}} = 20\,\mbox{meV}$. Inset shows dependence of the drag densities on the Rabi splitting at $L=10\,\mbox{nm}$.}
\end{figure}

In Fig. \ref{fig:p} we show the drag densities for 2D superconductors as functions of the interlayer distance $L$. Due to decay of the electron-polariton interaction with increasing $L$, $n_\mathrm{dr}$ decreases approximately as $L^{-1}$ (except NbSe$_2$). At the lowest distances $L\sim 5-10\,\mbox{nm}$, $n_\mathrm{dr}$ can reach $10^3-10^5\,\mbox{cm}^{-2}$. At moderate distances, $n_\mathrm{dr}$ is the highest for FeSe, while NbSe$_2$ demonstrates better results at lower Rabi splittings $\Omega_\mathrm{R}$, as seen in the inset. The large-gap superconductors FeSe, YBCO, and LSCO show increasing trends $n_\mathrm{dr}$ vs. $\Omega_\mathrm{R}$ due to the interplay between frequency dependencies of electron $\boldsymbol{\mathfrak{F}}^\mathrm{el}(\mathbf{q},i\omega_n)$ and polariton $\boldsymbol{\mathfrak{F}}^\mathrm{p}(\mathbf{q},i\omega_n)$ rectification functions. Namely, $\boldsymbol{\mathfrak{F}}^\mathrm{el}(\mathbf{q},i\omega_n)$ and $\boldsymbol{\mathfrak{F}}^\mathrm{p}(\mathbf{q},i\omega_n)$ have maxima at Matsubara frequencies equal to characteristic bosonic excitation energies: at $\omega_n\sim\Delta$ and $\omega_n\sim\Omega_\mathrm{R}$, respectively. With increasing $\Omega_\mathrm{R}$ their maxima come closer for large-gap superconductors, which increases the resulting integral (\ref{chi2}) and hence $n_\mathrm{dr}$ (\ref{n_dr}). In contrast, in NbSe$_2$ where $\Delta\ll\Omega_\mathrm{R}$, the drag density is almost independent on $\Omega_\mathrm{R}$.

\begin{figure}[t]
\begin{center}
\includegraphics[width=\columnwidth]{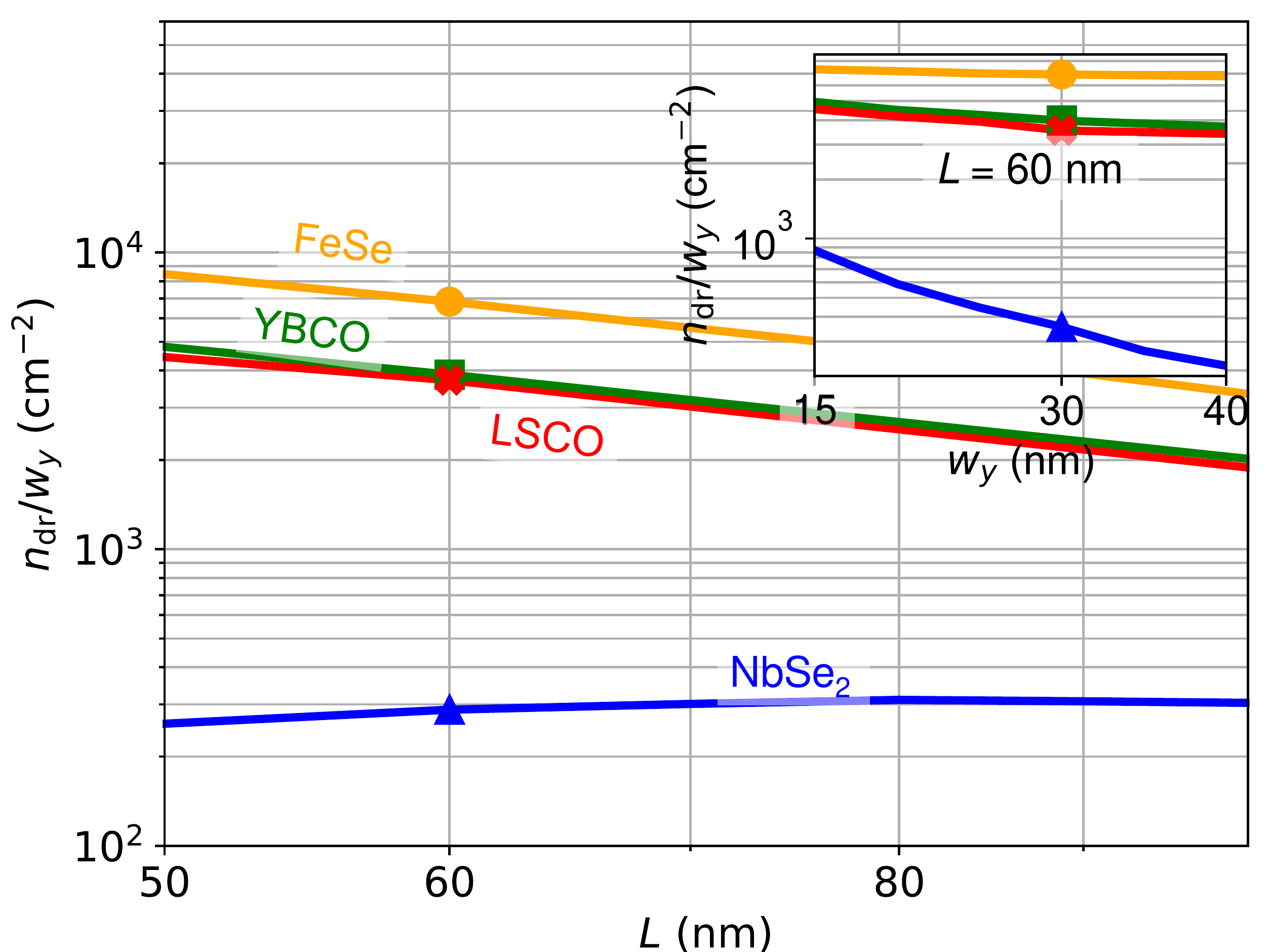}
\end{center}
\caption{\label{fig:w}Drag density $n_\mathrm{dr}$ as function of interlayer distance $L$ between Bose-condensed indirect polaritons and 1D nanoribbon superconductors. The Rabi splitting is $\Omega_{\mathrm{R}} = 20\,\mbox{meV}$, and the ribbon width is $w_y=30\,\mbox{nm}$ \cite{Marsili_2011}. Inset shows the dependence of $n_\mathrm{dr}$ on the nanoribbon width $w_y$ at $L=60\,\mbox{nm}$.}
\end{figure}

The drag density for atomically thin superconductors in the 1D nanoribbon geometry is shown in Fig.~\ref{fig:w}. To compare the 1D and 2D geometries, we divide $n_\mathrm{dr}$ by the ribbon width $w_y$, so that $n_\mathrm{dr}$ for 2D superconductors and $n_\mathrm{dr}/w_y$ for 1D nanoribbons or nanowires have the same dimensionality $\mbox{cm}^{-2}$. We find that $n_\mathrm{dr}/w_y$ can be of the same order $10^3-10^5\,\mbox{cm}^{-2}$ as in 2D geometry (Fig.~\ref{fig:p}) but at considerably larger interlayer distances $L$. This happens because of weaker metallic screening of the interlayer electron-polariton interaction [see Eq.~(\ref{epsilon1D})] by narrow superconductor ribbons. Note that for NbSe$_2$ $n_\mathrm{dr}$ even increases with $L$ because the weakening of the screening at increased interlayer separation outperforms decay of the bare electron-polariton interaction $V^\mathrm{el-p}(q)$. Inset in Fig.~\ref{fig:w} shows that $n_\mathrm{dr}$ weakly depends on $w_y$ when $w_y\ll L$ because of weak, approximately logarithmic dependence of the screening dielectric constant on the dimensionless parameter $qw_y\ll1$, where $q\sim L^{-1}$ is the characteristic transferred momentum [see (\ref{epsilon1D})]. When $w_y$ approaches $L$, our simplified analytical treatment of the screening in the coupled 2D-1D system becomes poorly applicable.

\begin{figure}[t]
\begin{center}
\includegraphics[width=\columnwidth]{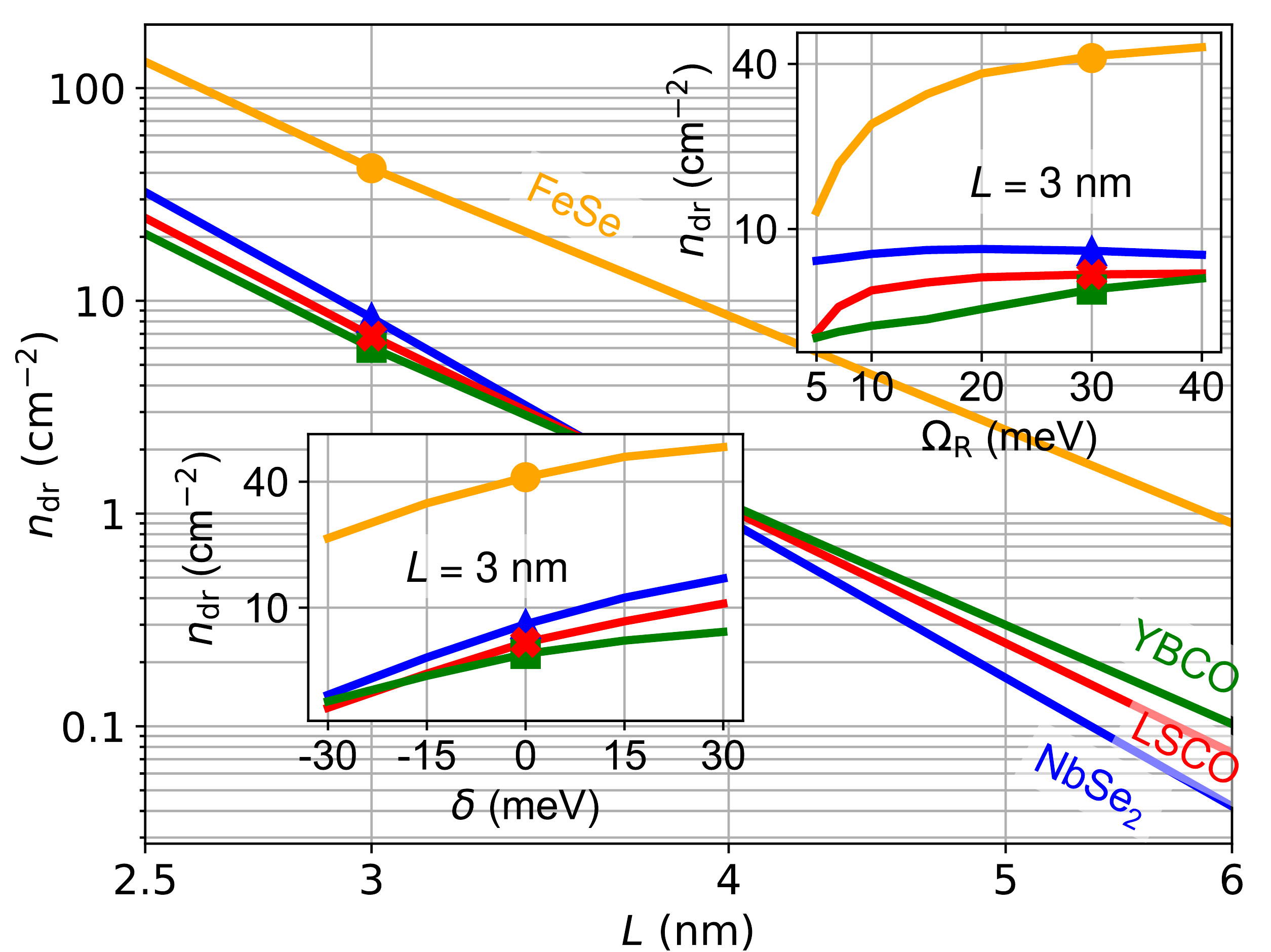}
\end{center}
\caption{\label{fig:dir2}Drag density $n_\mathrm{dr}$ as function of interlayer distance $L$ between Bose-condensed direct polaritons and 2D superconductors. The Rabi splitting on the main plot is $\Omega_\mathrm{R}=30\,\mbox{meV}$. Insets shows the dependence of the drag density on the nonzero photon-to-exciton detuning $\delta$ and on the Rabi splitting $\Omega_\mathrm{R}$ at $L=3\,\mbox{nm}$.}
\end{figure}

For the sake of comparison, we investigate the case of spatially direct polaritons, which interact with electrons via excitonic polarizability. This interaction is much weaker than with indirect polaritons \cite{Aminov_2022}, although with TMDC monolayer the microcavity setup becomes simpler and the achievable electron-to-exciton distances $L$ can be smaller. We consider the direct excitons hosted by TMDC monolayer with polarizability $30\,\mbox{nm}^3$ \cite{Garm} and Rabi splitting $\Omega_\mathrm{R}=30\,\mbox{meV}$ \cite{QFl}; the exciton-exciton interaction is $g^\mathrm{x-x}=0.1\,\mu\mbox{eV}\cdot\mu\mbox{m}^2$ \cite{Byrnes_2014}. The drag density for direct polaritons in 2D superconductor geometry is shown in Fig.~\ref{fig:dir2} as function of interlayer distance $L$. As seen, $n_\mathrm{dr}$ reaches $10-100\,\mbox{cm}^{-2}$ at smallest $L\sim2-3\,\mbox{nm}$, which is 2-3 order of magnitude lower than in the case of indirect polaritons. Moreover, the electron-polariton interaction in this case decays faster with increasing $L$, so $n_\mathrm{dr}\sim L^{-6.7}$ decreases much steeper as well. The second inset in Fig.~\ref{fig:dir2} shows that $n_\mathrm{dr}$ typically increases at positive photon-to-exciton detuning $\delta$ because the exciton polaritons become more excitonic-like.

\begin{figure}[t]
\centering
\includegraphics[width=1\columnwidth]{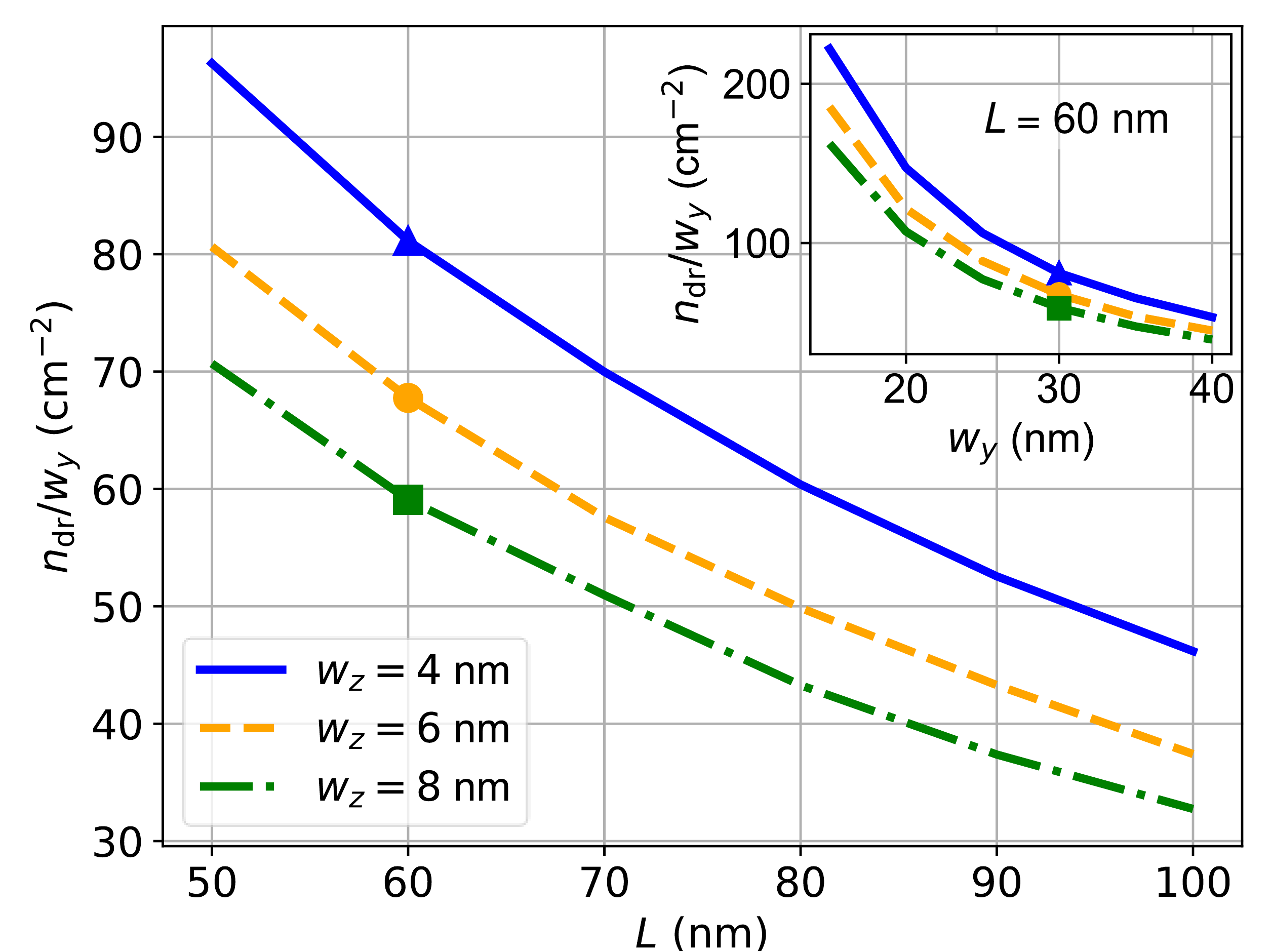}
\caption{Drag density $n_\mathrm{dr}$ as function of interlayer distance $L$ between between Bose-condensed indirect polaritons and 1D superconducting $\mathrm{Nb}_{1-x}\mathrm{Ti}_x\mathrm{N}$ nanowires of different thicknesses $w_z$. The wire width is $w_y=30\,\mbox{nm}$ and the Rabi splitting is $\Omega_\mathrm{R}=20\,\mbox{meV}$. Inset shows dependence of $n_\mathrm{dr}$ on the wire width $w_y$ at $L=60\,\mbox{nm}$, $w_z=6\,\mbox{nm}$.}
\label{fig:NbN}
\end{figure}

Finally, we consider the drag between indirect polaritons and 1D superconductor nanowires [Fig.~\ref{Fig1}(c)] fabricated from $\mathrm{Nb}_{1-x}\mathrm{Ti}_x\mathrm{N}$ bulk superconductors. The calculation results plotted in Fig.~\ref{fig:NbN} demonstrate that $n_\mathrm{dr}$ decreases approximately as $L^{-1.3}$ with increasing $L$. Generally  $n_\mathrm{dr}$ is two orders of magnitude smaller than in the case of 1D ribbons at considerable interlayer distances (compare with Fig.~\ref{fig:w}). The reason is the strong interaction screening due to high electron density in bulk superconductors. Fabrication of superconductor nanowires with smaller thicknesses $w_z$ allows to weaken the screening and increase $n_\mathrm{dr}$, although it poses additional challenges from the experimental point of view. As shown in the inset in Fig.~\ref{fig:NbN}, decreasing the wire width $w_y$ also allows to increase $n_\mathrm{dr}$.

\section{Discussion}\label{Sec5}

Experimental detection of the non-dissipative drag, as suggested in \cite{Aminov_2022}, can rely on accelerating polariton Bose condensate up to the velocity $v_\mathrm{p}$ and measuring of induced drag current in the superconductor. Electric current $J_\mathrm{el}$ flowing through 2D superconductor of transverse width $w_y$ is related to the electron current density (\ref{n_dr}) as $J_\mathrm{el}=ew_yj_\mathrm{el}$. The resulting estimate of drag-induced current is
\begin{align}
J_\mathrm{el}(\mbox{nA})&=1.6\times10^{-8}\nonumber\\
&\times w_y(\mbox{nm})\times n_\mathrm{dr}(\mbox{cm}^{-2})\times v_\mathrm{p}(10^9\,\mbox{cm/s}).\label{J_el}
\end{align}
If the polariton velocity is $v_\mathrm{p}=10^9\,\mbox{cm/s}$ \cite{Lerario2017}, the drag density reaches the highest value $n_\mathrm{dr}=10^5\,\mbox{cm}^{-2}$ reported in Fig.~\ref{fig:p}, and the width of 2D superconductor is $w_y=10^3\,\mbox{nm}$, we obtain $J_\mathrm{el}\sim2\,\mbox{nA}$. For 1D ribbon or wire superconductor the formula (\ref{J_el}) can be used with the replacement $n_\mathrm{dr}\rightarrow n_\mathrm{dr}/w_y$, and for the maximal ratio $n_\mathrm{dr}/w_y=5\times10^4\,\mbox{cm}^{-2}$ reported in Fig.~\ref{fig:w} we obtain $J_\mathrm{el}\sim0.02\,\mbox{nA}$. Collecting the drag-induced current from about 100 parallel wires would allow to raise the total current to the same order of 2 nA. This current is weak, but can be measured in accurate experiments \cite{Stehno_2016}. We can compare it with critical currents in the superconductors presented in Table~\ref{table:1}: multiplying 3D critical current density $j_\mathrm{c}^\mathrm{3D}$ by the superconductor film thickness $h$ and by its typical transverse size $w_y=10^3\,\mbox{nm}$ (in the case of 2D geometry) we obtain critical currents $J_\mathrm{c}=hw_yj_\mathrm{c}^\mathrm{3D}$ in the range $10^{-4}-10^{-6}\,\mbox{A}$, i.e. 3-5 orders of magnitude stronger than $J_\mathrm{el}$. On the other hand, typical critical currents of Josephson nanojunctions are tens of nA \cite{Stehno_2016}, so these structures can provide sufficient sensitivity to detect the drag-induced current.

Besides two-dimensional TMDCs chosen in this paper as a host for direct and indirect excitons, we can consider conventional semiconducting GaAs-based quantum wells as a possible candidate. However such systems pose two major problems. First, large Bohr radius of excitons $a_\mathrm{B}\gtrsim10\,\mbox{nm}$ \cite{Korti_Baghdadli_2013} imposes the limit on the maximum polaritonic density $n_\mathrm{p}$. If we restrict $n_\mathrm{p}$ to be 1-2 orders lower than the Mott critical density $\sim a_\mathrm{B}^{-2}$, we obtain $n_\mathrm{p}\lesssim10^{10}\,\mbox{cm}^{-2}$. It is 2 orders lower than polariton densities achievable in TMDCs \cite{Lerario2017}. Since the polaritonic rectification function $\boldsymbol{\mathfrak{F}}^\mathrm{p}$ is approximately proportional to $n_\mathrm{p}$ [see Eq.~(\ref{F_p2})], the drag density achievable in the quantum-well based systems is correspondingly lower. Second, the thickness of quantum wells ($h\gtrsim8\,\mbox{nm}$ \cite{High2012}) is much larger than that of atomically thin TMDCs ($h\sim1\,\mbox{nm}$), as well as the interwell distance, which is $d=8$ nm \cite{High2012} for quantum wells and $d=4$ nm for TMDCs; this imposes the lower limit on the interlayer distance $L$. For indirect excitons the minimum distance is $L\approx h+d/2$, which yields 12 nm and 3 nm for quantum well and TMDCs respectively. Since the drag density rapidly decreases with increasing $L$ as $L^{-2}-L^{-5}$ \cite{Aminov_2022}, in the former case it is expected to be 1-3 orders of magnitude lower even at the same polariton density.

To conclude, we extended the theory of non-dissipative drag \cite{Aminov_2022} on systems with non-quadratic electron and polariton dispersions, d-wave superconductors, and nanoribbon or nanowire superconductor geometries. The important point of our approach is taking into account dynamical screening of electron-polariton interaction by both polariton and electron subsystems. We carried out calculations of drag density in realistic conditions between exciton-polariton Bose condensate and atomically thin superconductors $\mathrm{NbSe}_2$, FeSe, YBCO, LSCO, as well as for thin-film bulk superconductors $\mathrm{Nb}_{1-x}\mathrm{Ti}_x\mathrm{N}$. The choice of atomically thin or nanowire superconductors is dictated by necessity to minimize light absorption in the microcavity thereby preserving its high quality factor. Interestingly, a large superconducting gap by itself is not essential for the non-dissipative drag between polaritons and superconductor. 

In realistic conditions, the drag density reaches $10^3-10^5\,\mbox{cm}^{-2}$ in the case of atomically thin superconductors $\mathrm{NbSe}_2$, FeSe, YBCO, and LSCO interacting with spatially indirect dipolar polaritons. Similar values of $n_\mathrm{dr}/w_y$ are achieved in the case of nanoribbons of the same materials with the width $w_y\sim30\,\mbox{nm}$, but at larger electron-to-exciton distances $L$. The resulting drag-induced superconducting currents can reach 2~nA, which are rather weak, but can be measured in accurate experiments. The main limiting factor for the drag is the screening of electron-polariton interaction by the metallic-like superconductor. In the case of thin-film bulk superconductors, such as $\mathrm{Nb}_{1-x}\mathrm{Ti}_x\mathrm{N}$, the drag is 2-3 orders of magnitude weaker even in nanowire geometry. Similarly, the drag is suppressed in the case of direct polaritons, which lack the persistent dipole moment and thus interact with electrons of the superconductor much weaker.

The drag density generally increases with increase of the Rabi splitting because the characteristic energies of excitations (Bogoliubov quasiparticles in the polariton subsystem and broken Cooper pairs in superconductors) come closer to each other. Also drag density increases at positive photon-to-exciton detunings, which make the polaritons more excitonic-like. Other ways to enhance the non-dissipative drag may include using the resonant effects or alternative system geometries, such as those based on optical-fiber \cite{Sedov2020} or nanocavity \cite{Vuckovic_2017} photon resonators.

\section*{Acknowledgments}
The work was supported by the Russian Foundation for Basic Research (RFBR) within the Project No. 21–52–12038. The work on analytical calculation of the superfluid drag density was done as a part of the research project FFUU-2021-0003 of the Institute for Spectroscopy of the Russian Academy of Sciences. The work on numerical calculations was supported by the Program of Basic Research of the
Higher School of Economics.

\appendix

\section{Rectification functions}\label{Appendix_A}

The electron and polariton nonlinear rectification functions entering Eq.~(\ref{chi2}) are defined as
\begin{align}
\boldsymbol{\mathfrak{F}}^\mathrm{el}(\mathbf{q},i\omega_n)&=\frac1A\int\limits_0^{1/T}d\tau_1 \int\limits_0^{1/T}d\tau_2\:e^{-i\omega_n(\tau_1-\tau_2)}\nonumber\\
&\times\left\langle T_\tau\mathbf{j}_0^\mathrm{el}(0)n^\mathrm{el}_\mathbf{q}(\tau_1)n^\mathrm{el}_{-\mathbf{q}}(\tau_2)\right\rangle,\label{F_el1}\\
\boldsymbol{\mathfrak{F}}^\mathrm{p}(\mathbf{q},i\omega_n)&=\frac1S\int\limits_0^{1/T}d\tau_1 \int\limits_0^{1/T}d\tau_2\:e^{-i\omega_n(\tau_1-\tau_2)}\nonumber\\
&\times\left\langle T_\tau\mathbf{j}_0^\mathrm{p}(0)n^\mathrm{p}_\mathbf{q}(\tau_1)n^\mathrm{p}_{-\mathbf{q}}(\tau_2)\right\rangle.\label{F_p1}
\end{align}
Here the Fourier harmonics of electron and polariton density operators are $n^{\mathrm{el}}_\mathbf{q}=\sum_{\mathbf{k}s} a^\dag_{\mathbf{k}s}a_{\mathbf{k}+\mathbf{q},s}$, $n^{\mathrm{p}}_\mathbf{q}=\sum_\mathbf{k}b^\dag_\mathbf{k}b_{\mathbf{k}+\mathbf{q}}$, and $S$ is the area of excitonic layer. In the clean limit, the electron rectification function (\ref{F_el1}) is given by the one-loop expression \cite{Aminov_2022}:
\begin{align}
\boldsymbol{\mathfrak{F}}^\mathrm{el}(\mathbf{q},i\omega_n)&=\frac{g_\mathrm{v}T}{A\hbar}\sum_{\mathbf{k}\epsilon_m} \frac{d \epsilon^\mathrm{el}_\mathbf{k}}{d\mathbf{k}}\,\mathrm{Tr}\left[\sigma_z\hat{G}^\mathrm{el}(\mathbf{k},i\epsilon_{m})\right.\nonumber\\
&\times\left.\hat{G}^\mathrm{el}(\mathbf{k},i\epsilon_{m})\sigma_z \hat{G}^\mathrm{el}(\mathbf{k}+\mathbf{q},i\epsilon_{m}+i\omega_{n})\right]\nonumber\\
&+(\mathbf{q},i\omega_n \rightarrow-\mathbf{q},-i\omega_n),\label{F_el2}
\end{align}
where $\epsilon_m=\pi T(2m+1)$ are fermionic Matsubara frequencies, $g_\mathrm{v}$ is the valley degeneracy factor which equals to 3 for NbSe${}_2$ and 1 for other materials from Table~\ref{table:1}. The electron Green functions $\hat{G}^\mathrm{el}$ are written in the Nambu representation as $(2\times2)$ matrices; they are found from the Gor'kov equations. 

For 2D copper oxide superconductors, where the gap is rather large and anisotropic, and the Fermi surface is strongly non-circular, we take the tight-binding anisotropic electron dispersion
$\epsilon^\mathrm{el}_\mathbf{k}=-2t(\cos k_xa+\cos k_ya)-4t'\cos k_xa\cos k_ya$, where $t=300\,\mbox{meV}$, $a=0.38\,\mbox{nm}$ or $a=0.53\,\mbox{nm}$, and $t'/t=-0.1$ or $-0.3$ for LSCO or YBCO respectively \cite{Yang_2006}. The anisotropic momentum-dependent gap is $\Delta_\mathbf{k}=\Delta(\cos k_xa-\cos k_ya)$.

Dimensionality of vectors in Eq.~(\ref{F_el2}) depends on the superconductor geometry. For 2D thin-film geometry [Fig.~\ref{Fig1}(a)], the rectification function $\boldsymbol{\mathfrak{F}}^\mathrm{el}$, electron momentum $\mathbf{k}$, and transferred momentum $\mathbf{q}$ are 2D vectors lying in the $(x,y)$ plane. For nanoribbon superconductor [Fig.~\ref{Fig1}(b)], $\boldsymbol{\mathfrak{F}}^\mathrm{el}$ and $\mathbf{q}$ are 1D vectors directed along the $x$ axis, while the electron momentum $\mathbf{k}$ is 2D vector. For relatively thick nanowire-shaped superconductor [Fig.~\ref{Fig1}(c)], $\boldsymbol{\mathfrak{F}}^\mathrm{el}$ and $\mathbf{q}$ are 1D vectors directed along the $x$ axis, while electron momentum $\mathbf{k}$ is 3D vector. In the last two cases, only the $x$-projection of the electron group velocity $d \epsilon^\mathrm{el}_\mathbf{k}/d\mathbf{k}$ is retained in Eq.~(\ref{F_el2}).

As shown in \cite{Aminov_2022}, in the limit $T=0$, $\Delta \ll E_\mathrm{F},\Omega_\mathrm{R}$ only the electron Fermi surface contributes to the momentum sum in (\ref{F_el2}), and for isotropic gap we obtain
\begin{equation}
\boldsymbol{\mathfrak{F}}^\mathrm{el}(\mathbf{q},i\omega_n)\approx \frac{4ig_\mathrm{v}}{A\hbar}\mathrm{Im}\sum_{\mathbf{k}}\frac{d \epsilon^\mathrm{el}_\mathbf{k}}{d\mathbf{k}}\,\frac{\delta\left(\epsilon^\mathrm{el}_\mathbf{k}\right)}{i\omega_n+\epsilon^\mathrm{el}_\mathbf{k}-\epsilon^\mathrm{el}_{\mathbf{k}+\mathbf{q}}}.\label{F_el3}
\end{equation}
When electrons have well-defined effective mass $m_\mathrm{el}^*$, so their Fermi surface is circular or spherical with the radius $k_\mathrm{F}$, we obtain analytical expressions:
\begin{equation}
\boldsymbol{\mathfrak{F}}^\mathrm{el} (\mathbf{q},i\omega_n)=2i\hat{\mathbf{q}}\frac{g_\mathrm{v}m_\mathrm{el}^*}{\pi q\hbar^3}\,\mathrm{Im}\frac{b}{\sqrt{b^2-1}}\label{Fel2D_1} 
\end{equation}
for 2D superconductors,
\begin{equation}
\boldsymbol{\mathfrak{F}}^\mathrm{el} (\mathbf{q},i\omega_n)=2i\hat{\mathbf{q}} \frac{g_\mathrm{v}m_\mathrm{el}^*w_y}{\pi q\hbar^3}\,\mathrm{Im}\frac{b}{\sqrt{b^2-1}}\label{Fel2D_2} 
\end{equation}
for 1D nanoribbon superconductors, and
\begin{align}
\boldsymbol{\mathfrak{F}}^\mathrm{el} (\mathbf{q},i\omega_n)&=2i\hat{\mathbf{q}} \frac{m_{\mathrm{e}}^{*}k_\mathrm{F}w_yw_z}{\pi^2q\hbar^3}\nonumber\\ &\times\mathrm{Im}\left[b\,\mathrm{arcsinh}\,\frac1{\sqrt{b^2 - 1}}\right]\label{Fel1D_3}
\end{align}
for 1D nanowire superconductors; here the dimensionless combination $b=(i\omega_nm^*_\mathrm{el} /\hbar^2-q^2/2)/qk_\mathrm{F}$ is introduced.

The rectification function of the polariton system (\ref{F_p1}) is calculated in the Bogoliubov approximation \cite{Aminov_2022} with taking into account the dominating contribution of the condensate processes:
\begin{equation}
\boldsymbol{\mathfrak{F}}^{\mathrm{p}}(\mathbf{q},i\omega_n)=\frac1\hbar\frac{d\epsilon^{\mathrm{p}}_{\mathbf{q}}}{d\mathbf{q}} \frac{4i\omega_nn_\mathrm{p}\tilde\epsilon^\mathrm{p}_\mathbf{q}}{\left[(i\omega_n)^{2}-\left(E^{\mathrm{p}}_{\mathbf{q}} \right)^{2} \right]^{2}},\label{F_p2}
\end{equation}
where $n_\mathrm{p}$ is the polariton density which is assumed to be almost equal to that of polariton condensate, $\tilde\epsilon^\mathrm{p}_\mathbf{q}=\epsilon^{\mathrm{p}}_{\mathbf{q}}+c_{q}-c_{0} $ is the polariton dispersion $\epsilon^{\mathrm{p}}_{\mathbf{q}}$ renormalized by the interaction-induced self-energy $c_q =X_{q}^{2} X_0^2n_{\mathrm{p}}g^{\mathrm{x-x}}$ which is momentum-dependent due to the Hopfield coefficients $X_q$. The chemical potential $c_{0}$ is subtracted in $\tilde\epsilon^\mathrm{p}_\mathbf{q}$ to make the Bogoliubov quasi-particle dispersion $E^{\mathrm{p}}_{\mathbf{q}} = \sqrt{\tilde{\epsilon}^{\mathrm{p}}_{\mathbf{q}}(\tilde{\epsilon}^{\mathrm{p}}_{\mathbf{q}}+2c_{q})}$ gapless. The interaction between excitons $g^{\mathrm{x-x}}$ is assumed to be independent of momentum.

\section{Interaction screening}\label{Appendix_B}

In the case of 2D thin-film superconductor geometry, the dielectric function entering Eq.~(\ref{V_scr}) is \cite{Boev2019,Aminov_2022}
\begin{align}
\tilde\varepsilon(q,i\omega_n)&=1-\Pi^\mathrm{el}_\mathrm{2D}(q,i\omega_n)\nonumber\\
&\times\left\{V^\mathrm{el-el}_\mathrm{2D}(q)+[V^{\mathrm{el-p}}(q)]^2\tilde\Pi^\mathrm{p}(q,i\omega_n)\right\},\label{epsilon2D}
\end{align}
where $V^\mathrm{el-el}_\mathrm{2D}(q)=2\pi e^2/\varepsilon_\mathrm{env}q$ is the 2D Fourier transform of electron-electron Coulomb interaction screened with the dielectric constant $\varepsilon_\mathrm{env}$ of microcavity environment, $\Pi^\mathrm{el}_\mathrm{2D}(q,i\omega_n)$ is the density response (or polarization) function of 2D superconductor, and
\begin{equation}
\tilde\Pi^\mathrm{p}(q,i\omega_n)  = \frac{2n_\mathrm{p}\tilde{\epsilon}^{\mathrm{p}}_{\mathbf{q}}}{(i\omega_n)^{2}-(E^{\mathrm{p}}_\mathbf{q})^2 }
\end{equation}
is the density response function of interacting polaritons calculated in the Bogoliubov approximation \cite{Boev2019,Aminov_2022}.

In the case of 1D superconductor ribbon or wire, we need to take into account its density response $\Pi^\mathrm{el}_\mathrm{1D}(q_x,i\omega_n)$ only on the $x$-component of the transferred momentum $\mathbf{q}$, while the polariton system responds both to $q_x$ and to $q_y$, so the dielectric function is given by the expression
\begin{align}
&\tilde\varepsilon(q_x,i\omega_n)=1-\Pi^\mathrm{el}_\mathrm{1D}(q_x,i\omega_n)\nonumber\\
&\times\left\{V^\mathrm{el-el}_\mathrm{1D}(q_x)+\int\frac{dq_y'}{2\pi}[V^{\mathrm{el-p}}(q')]^2 \tilde\Pi^\mathrm{p}(q',i\omega_n)\right\},\label{epsilon1D}
\end{align}
where intermediate 2D momenta under the integral are $\mathbf{q}'=\{q_x,q_y'\}$. In the 1D Fourier transform of Coulomb interaction $V^\mathrm{el-el}_\mathrm{1D}(q_x)=2e^{2}K_{0}(q_xw)/\varepsilon_\mathrm{env}$ (where $K_0$ is the modified Bessel function of the second kind), we need to impose the lower cutoff $w$ to interelectron distance, which is taken to be $w=w_y$ in the case of ribbon and $w=\sqrt{w_yw_z}$ in the case of thick wire. Note that the expression (\ref{epsilon1D}) is applicable at $w_y,w_z\ll L$, when the ribbon or wire is thin with respect to the interlayer distance.

We approximate $\Pi^\mathrm{el}_\mathrm{2D}(q,i\omega_n)$ by the polarization function of a normal electron system instead of that of a superconductor \cite{Gabovich1973, Rickayzen1959, Anderson1958}, because these functions are almost equal in the range of momenta and frequencies $\omega_n\sim\Omega_{\mathrm{R}}, q\sim L^{-1}$, which provide the dominating contribution to the drag density. Polarization function of 2D normal conductor with circular Fermi surface and well-defined effective mass calculated in the random-phase approximation is \cite{Stern1967}
\begin{align}
\Pi_\mathrm{2D}^\mathrm{el}(q,i\omega_n)&=-2g_\mathrm{v}\mathcal{N}_\mathrm{2D}\left\{
\vphantom{\sqrt{1-\left(\frac{q}{2p_{\mathrm{F}}}-i\frac{\omega}{qv_{\mathrm{F}}}\right)^{2}}}\frac12+\frac{ik_\mathrm{F}}q\,\mathrm{sign} \left(\omega_n\right)\right.\nonumber
\\ &\left.\times\sqrt{1-\left(\frac{q}{2k_\mathrm{F}}-\frac{i\omega_n m_{\mathrm{el}}^{*}}{qk_{\mathrm{F}}}\right)^{2}} \right\} + \mbox{c.c.},\label{Pi_2D}
\end{align}
where $\mathcal{N}_\mathrm{2D}= m^{*}_{\mathrm{el}}/2\pi\hbar^2$ is the 2D density of states at the Fermi level. For 2D copper oxide superconductors we calculate the density response function numerically as
\begin{align}
&\Pi^\mathrm{el}_\mathrm{2D}(q,i\omega_n)=\frac{T}{w_xw_y}\nonumber\\
&\times\sum_{\mathbf{k}\epsilon_m}\mathrm{Tr}\left[\hat{G}^\mathrm{el}(\mathbf{k},i\epsilon_m)\hat{G}^\mathrm{el}(\mathbf{k}+\mathbf{q},i\epsilon_m+i\omega_n)\right]\label{Pi_2D_RPA}
\end{align}
using the normal-state electron Green functions $\hat{G}^\mathrm{el}$ written in the Nambu representation. 

Density response function of 1D ribbon superconductor is obtained by taking $q_y=0$ in (\ref{Pi_2D}) and multiplying by its width $w_y$:
\begin{equation}
\Pi^\mathrm{el}_\mathrm{1D}(q_x,i\omega_n)=w_y\Pi^\mathrm{el}_\mathrm{2D}(q_x,i\omega_n).\label{Pi_1D_ribbon}
\end{equation}
For 1D wires we can similarly relate 1D and 3D density response functions as $\Pi^\mathrm{el}_\mathrm{1D}(q_x,i\omega_n)=w_yw_z\Pi^\mathrm{el}_\mathrm{3D}(q_x,i\omega_n)$. With the bulk superconductor $\mathrm{Nb}_{1-x}\mathrm{Ti}_x\mathrm{N}$ we investigate, we are working in the limit $q\ll k_\mathrm{F}$. Therefore the polarization function $\Pi^\mathrm{el}_\mathrm{3D}$ is dominated by close vicinity of the Fermi surface, so we can use the Lindhard function \cite{Mihaila_2011} taken in the long-wavelength limit
\begin{align}
\Pi^\mathrm{el}_\mathrm{3D}(q_x,i\omega_n)\approx- 2\mathcal{N}_\mathrm{3D}\left(1 +\frac{i\omega_n}{2v_\mathrm{F}q_x}\ln\frac{i\omega_n-v_\mathrm{F}q_x}{i\omega_n+v_\mathrm{F}q_x}\right),\label{Pi_1D_wire}
\end{align}
where $\mathcal{N}_\mathrm{3D}=m^{*}_{\mathrm{el}}  k_\mathrm{F}/2\pi^2\hbar^2$ is the 3D density of states at the Fermi level and $v_\mathrm{F}=\hbar k_\mathrm{F}/m_\mathrm{el}^*$ is the Fermi velocity.

\bibliography{References.bib}

\end{document}